\documentclass[aps,pra, twocolumn, superscriptaddress,amssymb,showpacs]{revtex4-1}
\usepackage{graphicx}
\usepackage{epstopdf}
\begin{document}
\title{Excitations of photon-number states in Kerr nonlinear  resonator at finite temperatures }
\author{G.~H.~ Hovsepyan}
\email[]{gor.hovsepyan@ysu.am}
\affiliation{Institute for Physical Researches,
National Academy of Sciences,\\Ashtarak-2, 0203, Ashtarak,
Armenia}
 
\author{G.~Yu.~Kryuchkyan}
\email[]{kryuchkyan@ysu.am}
\affiliation{Institute for Physical Researches,
National Academy of Sciences,\\Ashtarak-2, 0203, Ashtarak,
Armenia}\affiliation{Yerevan State University, Centre of Quantum Technologies and New Materials, Alex Manoogian 1, 0025,
Yerevan, Armenia}
\pacs{}
\begin{abstract}
We investigate temperature reservoir effects in a lossy Kerr nonlinear resonator considering  selective excitation of ooscillatory mode driven by a sequence of Gaussian pulses. In this way, we analyze time-dependent populations of photon-number states and quantum statistics on the base of the second-order photon correlation function in one-photon and two-photon transitions.   The effects coming from 
thermal reservoirs are interesting for performing more realistic approach to generate Fock states and for study of 
phenomena connecting quantum engineering and temperature. We also study the role of pulse-shape effects for selective excitation of oscillatory mode.
\end{abstract}
\maketitle
\section{Introduction}
\label{intro}
Anharmonic oscillators or resonators involving Kerr-type nonlinearities have been an
interesting research topic due to their broad applications
in technology and fundamental physics \cite{kov}, \cite{dyk}. 
Recently, this systems with strong parameters of the nonlinearity are also
attracting considerable attention for investigation of various quantum effects.  The quantum dynamics of the Kerr nonlinear resonator (KNR) is naturally described by Fock states, which have a defnite number of
energy quanta. For systems with strong Kerr nonlinearity, leading to the photon-photon interaction in QED systems, the nonlinearity makes frequencies of transitions between
adjacent oscillatory energy levels different, i.e. nonlinearity effects in an anharmonic oscillator break the equidistance
of oscillatory energy levels. Thus, in the case of strong nonlinearity the oscillatory energy levels are well resolved
and spectroscopic selective excitation of transitions between Fock states at the level of a few quanta becomes to be possible.  This regime has been demonstrated in the photon blockade (PB) \cite{Liu,A,Mir,gor,nori},  generation of Fock states and superposition states in KNRs \cite{nori,gev2,mod,sup} and on changing PB into electromagnetically-induced
transparency  \cite{nori1}. 
Note, that nonstationary PB in a context of Fock states generation  was predicted in \cite{schmid} for short-time evolution of  a kicked KNR and then in \cite{imam6}. The nonstationary Kerr PB is often referred to as a nonlinear optical-state truncation or nonlinear quantum scissors \cite{leon25}.

  The important parameter responsible for KNR to reach quantum regimes is the ratio between the parameter of the Kerr-type nonlinearity and damping of the oscillatory mode. Thus, efficiency of quantum
nonlinear effects requires a high nonlinearity with respect to dissipation.
 In this respect, strong nonlinearities on a few-photon level can be produced by interaction between photons and an atom in a cavity \cite{Birn2,Bish,Fink,Far}, in systems with interacting photons or polaritons in
arrays of cavities coupled to atoms or qubits \cite{Hart,Utus,Schm,Tom}, in optomechanical systems and Kerr type nonlinear cavities \cite{Lia,Ferr}. An important implementation of Kerr-type microwave resonator has been recently
achieved in the context of superconducting devices based
on the nonlinearity of the Josephson junction  revealing nonlinear behavior even
at the single-photon level   \cite{lang,hof,kir}.
  
Note, that most theoretical proposals on investigation of quantum effects in nonlinear Kerr resonators
are focused on using  idealized cases of zero temperature
reservoirs  since they can lead to the study of pure quantum effects.  However, consideration of  the reservoir at finite temperatures leads to applications in simulating of more realistic systems as well as to study  of unusual quantum
phenomena connecting quantum engineering and temperature.

In this paper we investigate finite temperature reservoir effects in the dissipative KNR considering  selective excitation in transitions between Fock states.  For this goal 
we investigate selective excitation  in the pulsed regime, considering KNR driven by a sequence
of classical Gaussian pulses separated by time intervals. This system has been used in the field of PB and generation of Fock states  \cite{gor,gev2,mod} as well as for production of superpositions states in a mesoscopic range  \cite{gev1} and for demonstration of chaos at a low level of quanta \cite{chaos1,chaos2}. 

  It has been demonstrated in the cited papers, that in the specific pulsed regime  the obtained results considerably differ from those derived for the case of continuous-wave  driving.  On the whole, the production of the Fock states 
as well as the superposition of the Fock states can be controlled by shape of pulses and is realized for time-intervals exceeding the characteristic
time of decoherence.  Thus, the other goal of the present paper is investigation of various  regimes of selective excitation in dependence from the parameters of pulses. In this way, we demonstrate   improving of the degree of quantum effects in KNR  by applying the sequence of tailored pulses. This approach was recently exploited for formation of high degree continuous-variable entanglement in the nondegenerate optical parametric
oscillator \cite{adam,adam2}

Our consideration is based on the Hamiltonian of KNR coupled with  finite temperature  reservoir and  the master
equation for the reduced density matrix. We focus on analysis of the mean photon number, the probability distributions of photons and the second-order correlation functions of photons for zero-delay time that describes quantum statistics of oscillatory mode in thermal Kerr resonator. It should be mentioned that  thermal reservoir effects have been widely  investigated in nonlinear optical processes, particularly, for the  resonance fluorescence in monochromatic field   \cite{res,kry,env} and for the  resonance fluorescence in bichromatic field  \cite{kry1,kry2,kry3}.

The paper is arranged as follows. In Sec. II we describe periodically pulsed KNR coupled with a finite temperature reservoir.  In Sec. III we consider one-photon and two-photon selective excitations on the base of  populations of photon-number states and the second-order correlation functions. In subsection 3.3 we also discuss dynamics of Fock states populations in dependence from both the duration of pulses and amplitude of pump field.  We summarize our results in Sec. IV.

\section{ Kerr nonlinear resonator coupled with  thermal reservoir}
In this section, we give the theoretical description of the
system. The Kerr nonlinear resonator driven by the field at the central  frequency $\omega$ and interacting with a reservoir
is described by the following Hamiltonian:
\begin{eqnarray}
H=\hbar \omega_{0}a^{+}a + \hbar \chi a^{+{2}}a^{2}+ ~~~~~~~~~~~~~~~~~~~~~~~~~~~~~~~\nonumber\\ 
  + \hbar f(t)({\Omega} e^{-i\omega t}a^{+} + {\Omega^{*}}e^{i\omega t}a)+H_{loss}.~~~~~
\label{H}
\end{eqnarray}
Here, time-dependent coupling constant $\Omega f(t)$  proportional to the  driving field amplitude consists of the Gaussian pulses with the duration $T$  separated by time intervals $\tau$
\begin{equation}
f(t)=\sum{e^{-(t - t_{0} - n\tau)^{2}/T^{2}}}, \label{driving}
\end{equation}
while $a^{+}$, $a$ are the creation and annihilation operators, $ \omega_{0} $  is an oscillatory frequency, $\chi$ is the
nonlinearity parameter.

$H_{loss}=a \Gamma^{+} + a^{+}\Gamma $
is responsible for the linear losses of oscillatory states, due to
couplings with heat reservoir operators giving rise to the
damping rate  $\gamma$. 
 We trace out the reservoir degrees of freedom in the Born-Markov limit assuming that system and environment are
uncorrelated at initial time t = 0. This procedure leads to the master
equation for the reduced density matrix. The master equation   within the
framework of the rotating-wave approximation, in the interaction
picture corresponding to the transformation $\rho  \rightarrow   e^{-i\omega  a^{+} a t}\rho e^{i\omega  a^{+} a t}  $ reads as 

\begin{equation}
\frac{d \rho}{dt} =-\frac{i}{\hbar}[H_{eff}, \rho] +
\sum_{i=1,2}\left( L_{i}\rho
L_{i}^{\dagger}-\frac{1}{2}L_{i}^{\dagger}L_{i}\rho-\frac{1}{2}\rho L_{i}^{\dagger}
L_{i}\right)\label{master},
\end{equation}
where $L_{1}=\sqrt{(n_{th}+1)\gamma}a$ and $L_{2}=\sqrt{n_{th}\gamma}a^+$ are the
Lindblad operators, $\gamma$ is a dissipation rate, and $n_{th}$ denotes the mean
number of quanta of a heat bath,  $n_{th}=\frac{1}{e^{\hbar \omega/ kT_{B} }-1}$. The effective Hamiltonian reads as

\begin{equation}
H_{eff}=\hbar \Delta a^{+}a + \hbar \chi a^{+2}a^{2} +
\hbar f(t)(\Omega a^{+} + \Omega^{*}a),\label{Hamiltonian}
\end{equation}
 where $\Delta=\omega_{0} -\omega$ is the detuning
between the mean frequency of the driving field and the frequency of the
oscillator.
 
 Note, that  analytical results for a dissipative driven nonlinear oscillator  in continius-wave,  steady-state regimes  have
been obtained in terms of the solution of the Fokker-Planck equation for the quasiprobability distribution function $P(\alpha, \alpha^{*})$ in complex
P-representation \cite{drum}. This approach based on the
method of potential equations leads to the analytic solution for the quasiprobability distribution function $P(\alpha,  \alpha^{*})$ within the framework
of an exact nonlinear treatment of quantum
fluctuations. In this way,  the probability distribution of
photon-number states and  Wigner functions  have been also obtained  \cite{a33,a34,kh}.

This model seems experimentally feasible and can be realized in
several physical systems. Particularly, the effective Hamiltonian (\ref{Hamiltonian}) (with $f(t)=1$ ) describes a
qubit off-resonantly coupled to a driven cavity. In fact, it is well known that the Hamiltonian of a two-level atom interacting with the cavity mode in the dispersive approximation, if the two-level system remains in its ground state, can be reduced to the effective Hamiltonian (\ref{Hamiltonian}). This model also describes a nanomechanical oscillator with $a^{\dagger}$ and $a$ raising and lowering operators related to the position and momentum operators of a mode of quantum motion. The other implementation is  a transmission-line resonator involved the  Josephson
junction with quadratic part of the Josephson potential. In this case  the $a^{+}$ and $a$ raising and lowering operators describe the normal mode of the resonator junction circuit \cite{bias}.
 Kerr-like systems seems to be important also  for production of  quantum correlation and maximally entangled stated. They were discussed, for instance,  in  \cite {leonski1,said,leonski2,said2}.

\section{Selective excitation of  Fock states}\label{SecPD}
In the absence of any driving, states of the Hamiltonian (\ref{Hamiltonian})
are the Fock photon states $|n\rangle$ which are spaced in the energies $E_{n} = E_{0} +
\hbar\omega_{0} n + \hbar\chi n(n-1)$ with $n = 0, 1, ...$. The
levels form an anharmonic ladder with
anharmonicity that is given by $E_{21}-E_{10}=2\hbar\chi$. For strong
nonlinearities 
$\chi/\gamma >1$ the nonlinear shifts of oscillatory energetic levels exceed the radiative widths of these levels. In this case the selective resonance excitations of Fock states in the transitions $|m\rangle\rightarrow|n\rangle$ are possible due to interaction with driving field. Considering the various selective excitations from vacuum state,  in the multiphoton transitions $|0\rangle\rightarrow|n\rangle$, we obtain the resonance n-photon transitions between oscillatory initial and final states on the frequencies  $n\omega_n=E_{n0}=n\omega_0+\chi n(n-1)$. Thus, the detuning for the resonant frequencies  $\Delta_{n}=\omega_{0} -\omega_{n}$ is calculated as $\Delta_{n}=-\chi(n-1)$. For one-photon transition, $n=1$, and $E_{10}=\hbar \omega_0$, the resonance frequency is $\omega_1=\omega_0$,
for two-photon transition, $n=2$,  and $E_{20}=2 \hbar \omega_0 + 2 \chi $, the  two-photon resonance frequency is $\omega_2=\omega_0+ \chi $, while for $n=3$ , $E_{30}=3 \hbar \omega_0 + 6 \chi $, the resonance frequency is $\omega_3=\omega_0+ 2\chi $.
Below, we concentrate on quantum regimes
for the parameters leading to resolved oscillatory energy levels calculating the mean photon number, the photon-number distributions and the second-order correlation functions  for photonic mode.

Considering the pulsed regimes of Kerr nonlinear reservoir we assume that the spectral widths of pulses should be smaller
than the nonlinear shifts of the oscillatory energy levels. It means that the
duration of pulses should be larger than $1/\chi $. Thus, for strong
nonlinearities 
$\chi/\gamma >1$, we arrive to the following inequalities for the duration of Gaussian pulses $1/\gamma > T > 1/\chi$.
Note, that the temporal pulse separation is 
larger than the cavity photon lifetime.

We solve the master equation Eq. (\ref{master}) numerically based on quantum
state diffusion method.  The applications of this method for studies of driven anharmonic oscillator 
can be found in \cite{gev2,mod,gor,chaos1}.
In the calculations, a finite basis of number states $|n\rangle$ is kept large
enough (where $n_{max}$ is typically 50) so that the highest energy states are
never populated appreciably. In the following the main photon number and the distributions of photon numbers
$P(n)=\langle n|\rho|n\rangle$ will be analyzed for various level of thermal photons. 
 We also turn to calculation of the normalized second-order correlation function for zero delay time $ g^{(2)}$  defined as:
\begin{equation}
g^{(2)}(t)=\frac{\langle a^{\dagger}(t)a^{\dagger}(t)a(t)a(t)\rangle}{(\langle a^{\dagger}(t)a(t)\rangle)^2}
\end{equation}
for observation of  photon statistics and photon correlation effects. Moreover, the correlation function is expressed through the mean photon number fluctuations as $\langle (\Delta n)^2\rangle = \langle n\rangle + \langle n\rangle ^2(g^{(2)}-1)$.
Thus, the condition $g^{(2)}<1$ corresponds to the sub-Poissonian statistics, $\langle (\Delta n)^2\rangle < \langle n\rangle$ .

In the pulsed regime these quantities  are nonstationary and exhibit a periodic time-dependent behavior, i.e. repeat the periodicity of the pump laser in an over transient regime.
Besides this, the results depend on the parameters of Gaussian pulses such as the amplitude, the duration of pulses and the time-interval between them, dissipation rates and Kerr-interaction coupling. Thus, we investigate production of photonic states for an arbitrary interaction time-intervals including also time-intervals
exceeding the characteristic time of dissipative processes, $t\gg\gamma^{-1}$. 
 In the cited paper  \cite{gor}, only the case of KNR with the mean number of reservoir photons $ N=0.001$ has been considered. In addition to this investigation, bellow we present systematic  analyse of  KNR for the  wide range of thermal photon numbers including also vacuum case, $N=0$.

\begin{figure}
\resizebox{0.5\textwidth}{7.5cm}{%
  \includegraphics{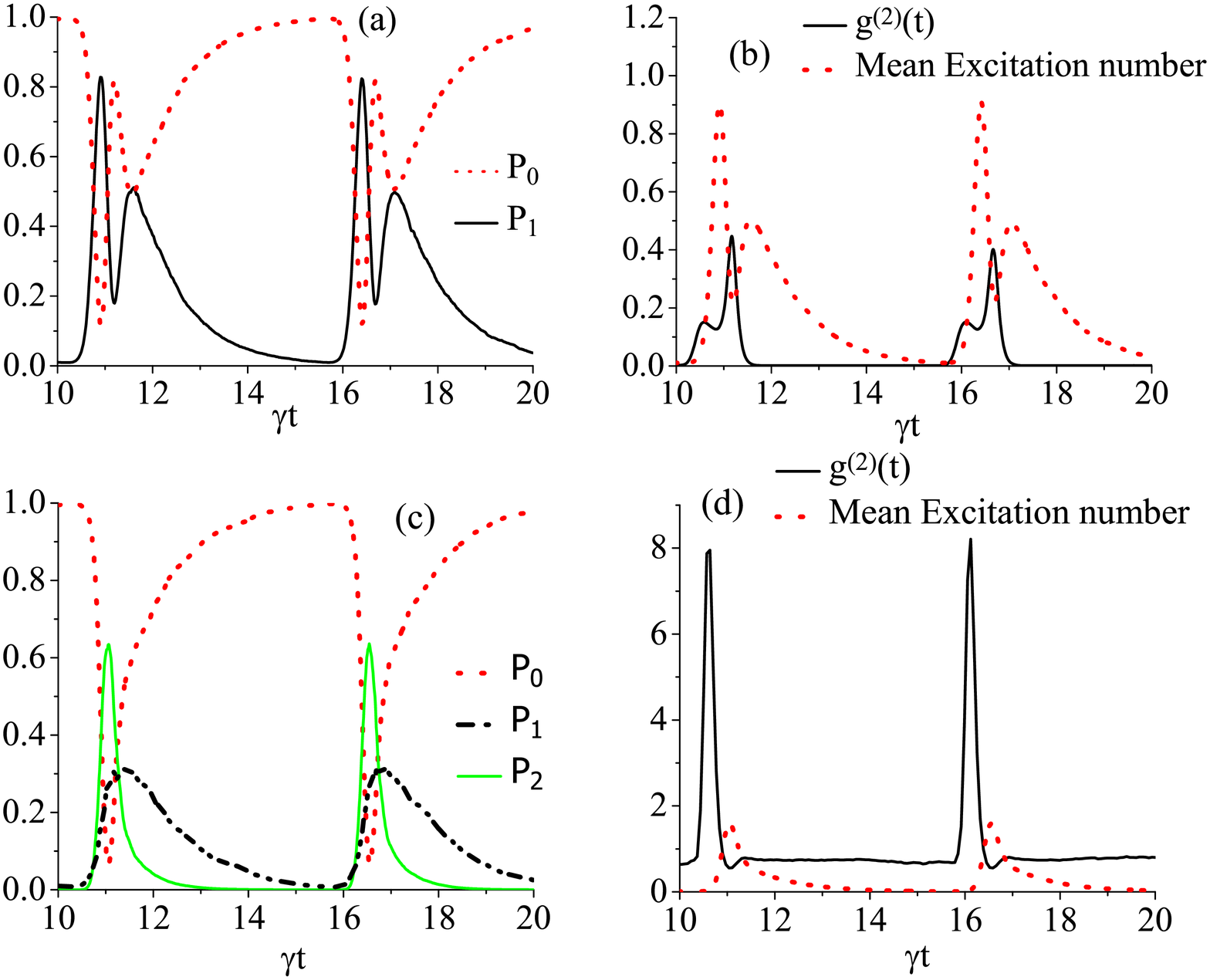}
}
\caption{The selective excitations in KNR at zero temperature for both cases of the detuning $\Delta_{1}=0$ (a), (b) and  $\Delta_{2}=\omega_{0} -\omega_{2}=-\chi$ (c), (d).  The populations of photon-number states  versus time-intervals (a), (c). The averaged photon numbers and the second-order correlation function versus time-intervals (b), (d). The parameters are: (a), (b)
$\chi /\gamma = 15$, the maximum amplitude of pump field $\Omega/\gamma =6$; (c), (d)
$\chi /\gamma = 30$, the maximum amplitude of pump field $\Omega/\gamma =12$ .  The mean number of reservoir photons $n_{th}=0$ and 
$\tau = 5.5{\gamma}^{-1}$, $T=0.4{\gamma}^{-1}.$}
\label{0}
\end{figure}

\subsection{The case of reservoir at zero temperature}

In this subsection, we consider the case of reservoir at zero temperature when only pure quantum effects occur. The typical results at the detuning $\Delta=\Delta_{1}=0$, i. e. at the frequency $\omega = \omega_{0}$,  corresponding to one-photon transition $|0\rangle\rightarrow|1\rangle$ are depicted in Figs.1 for  
$T = 0 K$.

 The populations
$P_0, P_1$ of the photon-number states $|n\rangle$ versus time-intervals is shown in Fig.\ref{0} (a), while the mean photon number and the correlation function are displayed in Fig.\ref{0} (b) for zero detuning $\Delta=0$. These results show the selective excitations of the Fock state $|1\rangle$. In this case  the excitation power is also small therefore the high excitation numbers have not been occupied. In the result only the $|1\rangle$ is effectively excitated. As we see, the population $ P_1$  reachs up to 0.8 at the maxima of the average number of photons.  The nonstationary 
correlation function versus dimensionless time intervals is shown in Fig.\ref{0} (b)
 with the curve of mean photon number. As we see,  the average number of photons at frequency $\omega_{0}$ follows almost the train of Gaussian pulses. The second small peak corresponds to Rabi oscillations and the  broadening of the shapes in comparisson with Gaussian envelopes are due to the radiative decays of the mode in a resonator. The mean photon number equals to $\langle n\rangle$=0.9 and  is approximatelly  presented as $\langle
n\rangle=P_1+2P_2$=0.9 in this regime. Thus, time-dependence of ${\langle n(t)\rangle}$ repeats the behaviour of the population  $ P_1$.

 During the pulses, if $P_1$ reachs to the maximum,  the probability of generation of
a second photon in  the mode  at the frequency $\omega = \omega_{0}$ is suppressed that is demonstrated in  the behavior of $g^{(2)}(t)$. In this case oscillatory mode acquires
sub-poissonian statistics with the second-order correlation function $ g^{(2)}<
1$.  As calculations show,
$g^{(2)}=0.12$ at the maximum values ${\langle n(t)\rangle}=0.9$ and is zero at the second small peak. This results are in accordance with the result that for the  pure $|1\rangle$ Fock
state the normalized second-order correlation function is zero.

\begin{figure}
\resizebox{0.5\textwidth}{7.5cm}{%
\includegraphics{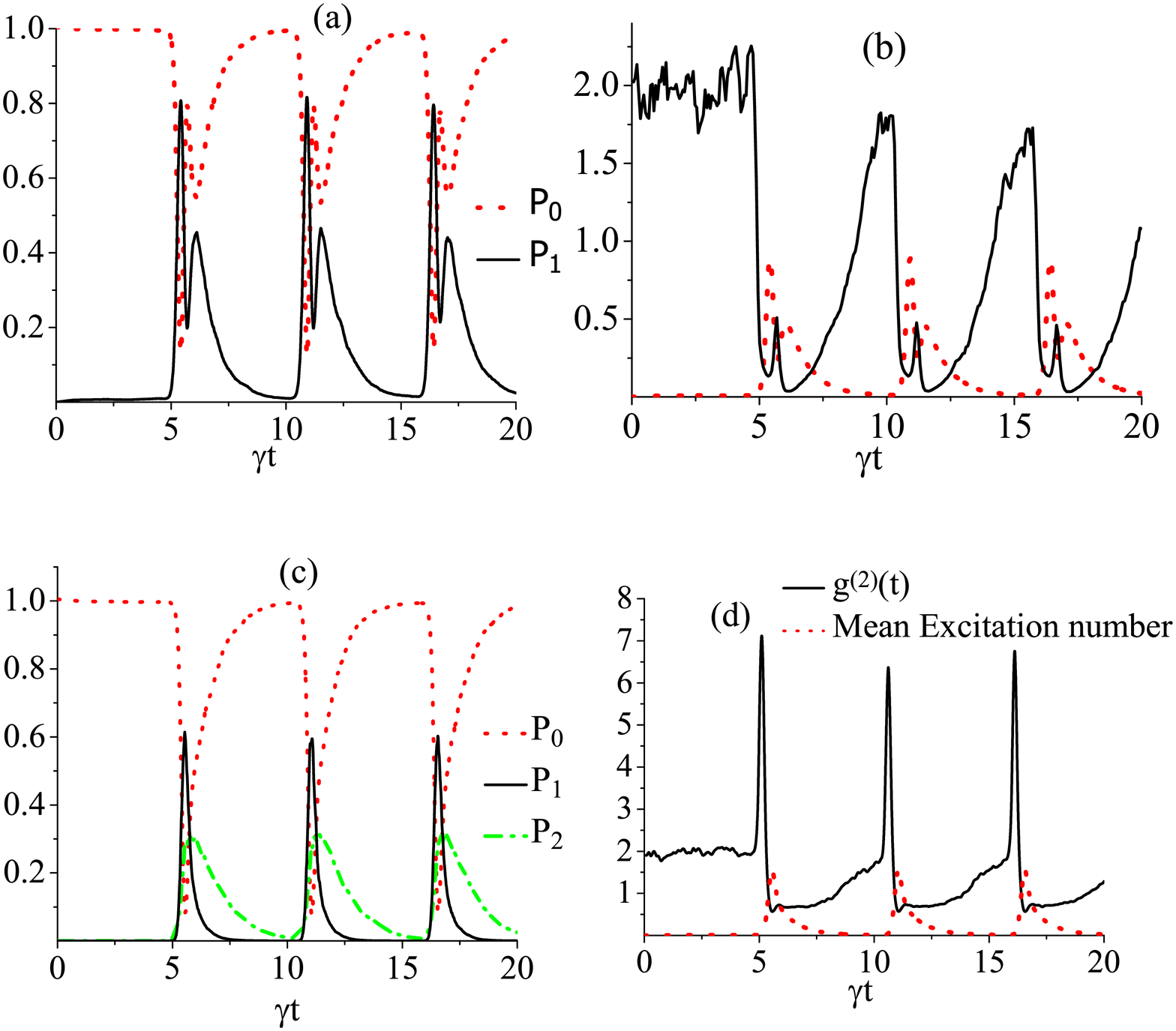}
}
\caption{Reservoir effects in the selective excitations in KNR for two cases of the detuning: $\Delta_{1}=0$ (a), (b) and  $\Delta_{2}=\omega_{0} -\omega_{2}=-\chi$ (c), (d).  The populations of photon-number states (a), (c). The averaged photon numbers and the second-order correlation function (b), (d). The parameters are: (a), (b)
$\chi /\gamma = 15$, the maximum amplitude of pump field $\Omega/\gamma =6$; (c), (d)
$\chi /\gamma = 30$, the maximum amplitude of pump field $\Omega/\gamma =12$ .  The mean number of reservoir photons $n_{th}=0.1$ and the parameters of pulses are:
$\tau = 5.5{\gamma}^{-1}$, $T=0.4{\gamma}^{-1}.$}
\label{01}
\end{figure}
\begin{figure}
\resizebox{0.5\textwidth}{7.5cm}{%
\includegraphics{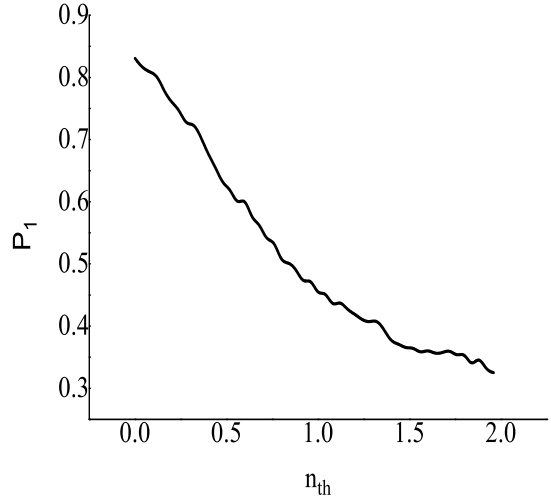}
}
\caption{The maximal values of population $P_1(t)$  versus $n_{th}$. The parameters are: $\Delta=0$,  $\chi /\gamma = 15$, the maximum amplitude of pump field $\Omega/\gamma =6$ , 
$\tau = 5.5{\gamma}^{-1}$, $T=0.4{\gamma}^{-1}.$}
\label{Stat}
\end{figure}

The results for the  other operational regime for zero-temperature reservoir are depicted in  Fig.1 (c),(d). Here, we assume 
 the regime of two-photon excitation of resonator mode, choosing $\Delta=\Delta_{2}=\omega_{0} -\omega_{2}$, at the frequency of resonance transition  $\omega_2=\omega_0+ \chi $.  Thus, in this case the detuning is $\Delta/\gamma=-\chi/\gamma$. 
These results indicate an effective  two-photon selective excitation as illustrated   in Fig.1 (c). As we see, 
 in this regime of two-photon excitation  the maximal population  $P_2=0.64$ exceeds  the maximal  population of one-photon state $P_1=0.3$ . The time evolution of the mean photon number versus
dimensionless time is depicted in Fig. \ref{0} (d)  with the plot of the second-order correlation function. In this case, the maximal value of mean photon numbers reachs to  ${\langle n(t)\rangle}=1.9$ while the level of photon-number correlation at this time interval is $g^{(2)}=0.6$ and the  variance of photon number fluctuations equals to $\langle (\Delta n)^2\rangle =0.24  \langle  n\rangle $, i.e. the oscillatory mode has sub-poissonian statistics.
The result on the second-order correlation function is in accordance with analytical result for pure state  $|2\rangle$. Indeed, it is easy to calculate  that $\langle 2|(a^{+})^2 a^2|2\rangle = 2$ and thus $g^{(2)}=0.5$ for  $|2\rangle$ Fock state.

\subsection{Reservoir effects in selective excitations of   Fock  states}

Considering the reservoir at finite temperatures we realize that temperature effects can lead  to decreasing   of the populations  in comparison to the case of zero-temperature reservoir on one side and change photon statistics of mode on the other side. The results for  thermal photons with $n_{th}=0.1$  are shown in Fig. \ref{01} (a),(c).  In order to illustrate the difference between the case of zero-temperature resonator  we  assume here the parameters
as in the previous case depicted in Fig.\ref{0} (a),(c). As we see, the populations only are slightly  decreased in the presence of temperature noise with $n_{th}=0.1$ for two considered regimes (a) and (c). We show in Fig. \ref{01} (b), (d) how the mean photon numbers and the second-order correlation function explicitly depend on the time-interval. As it can be seen, the maximal values of mean photon numbers are approximately the same as in the case of pure resonator, while quantum statistics of oscillatory mode is considerably changed  due to the thermal noise. Indeed, in this case  initial time-evolution of the system  until a time-interval corresponding to the first pulse coming to cavity is described by the master equation without the interaction part, $\Omega f(t)=0$.  Thus, in this range the quantity $g^{(2)}$ describes the   statistics of mode of an anharmonic oscillator in thermal reservoir. It displays  time-dependent fluctuations around the level  $g^{(2)}=2$ that correspond to the statitics of thermal light mode.  From the results in Fig. \ref{01} (b), (d) it can also be infered that the correlation function is sharply increasing in the fronts of pulses. However, the peak values strongly depend on the parameters of nonlinearity and amplitude of pump field. So, for Fig.\ref{01} (b) the maximal value is observed as $g^{(2)}=0.45$, while for  Fig.\ref{01}(d) the correation function equals to $g^{(2)}=7$, (for   ${\langle n(t)\rangle}=0.5$),   and corresponds to super-poissonan statistic of mode,  $\langle (\Delta n)^2\rangle =4 \langle n\rangle $ at this time-interval. 
\begin{figure}
\resizebox{0.5\textwidth}{7.5cm}{%
\includegraphics{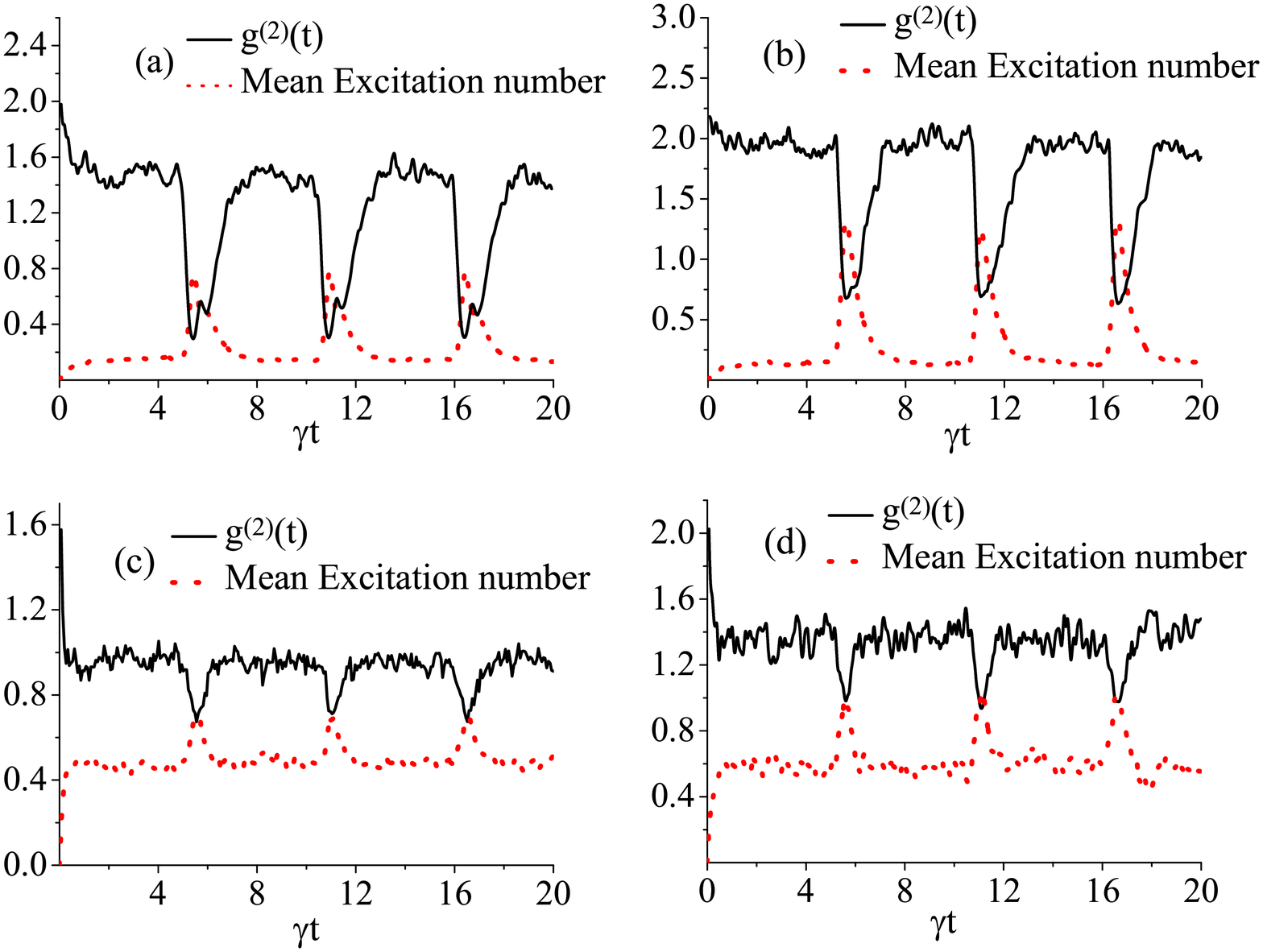}
}
\caption{Time-evolution of the averaged photon number and the second-order correlation function for two regimes of KNR. The parameters are: (a),(c) $\Delta=0$,
$\chi /\gamma = 15$, $\Omega/\gamma =6$;  (b),(d) $\Delta_{2}=\omega_{0} -\omega_{2}=-\chi$,
$\chi /\gamma = 30$,  $\Omega/\gamma =12$. The mean number of reservoir photons: (a),(b) $n_{th}=0.58$; (c),(d) $n_{th}=1.9$. The other parameters are:
$\tau = 5.5{\gamma}^{-1}$, $T=0.4{\gamma}^{-1}.$}
\label{05819}
\end{figure}

It is interesting to analyse excitations of resonator mode for high levels of  thermal photon numbers. In this way, in Fig. \ref{Stat} we plot the maximal values of the population $P_1$  in dependence on $n_{th}$ for the operational regime of KNR depicted in  Fig.\ref{0} and Fig.\ref{01}.  As we see the population $P_1$  is strongly  decreasing with increasing of the thermal photon level. 

The correlation functions  and the mean photon numbers for high  levels of thermal photons are considered in details for two cases, with  $n_{th}=0.58$ and $n_{th}=1.9$,  in Fig. \ref{05819}. We analyse finite temperature reservoir effects for one-photon and two-photon excitation regimes, corresponding to two values of the detuning:  $\Delta_{1}=0$  and  $\Delta_{2}=\omega_{0} -\omega_{2}=-\chi/\gamma$. 
As we see, in both these cases $g^{(2)}$ describes mainly the statistics of  dissipative oscillatory mode with the dips that correspond to the peaks of the averaged photon numbers. Particularly, for the regime depicted in  Fig. \ref{05819} (a) the peak of mean photon number $n=0.8$  while the corresponding dip on $g^{(2)}=0.3$ shows subpoissonian staistics of mode, $\langle (\Delta n)^2\rangle =0.56 \langle  n\rangle $.  The location of these dips and peaks in Fig. \ref{05819} are determined by time-intervals of Gaussian pulses. We also found that these dips and peaks  deacrese, if the level of thermal noise increases.

\subsection{Pulse-parameter effects in dynamics of the populations}

It should be mentioned that the  parameters of the Gaussian pulses in above consideration are  free parameters and they might be chosen in order to control pulsed selective excitation of resonator mode.  In the end of this section we shortly  illustrate  behavior of the populations by increasing  the duration of pulses on one side and by increasing the amplitude of driving field on the other side for zero-temperature reservoir. In this way, we describe  two regimes of excitation: one-photon and two-photon excitations of oscillatory mode.

\begin{figure}
\resizebox{0.5\textwidth}{4cm}{%
\includegraphics{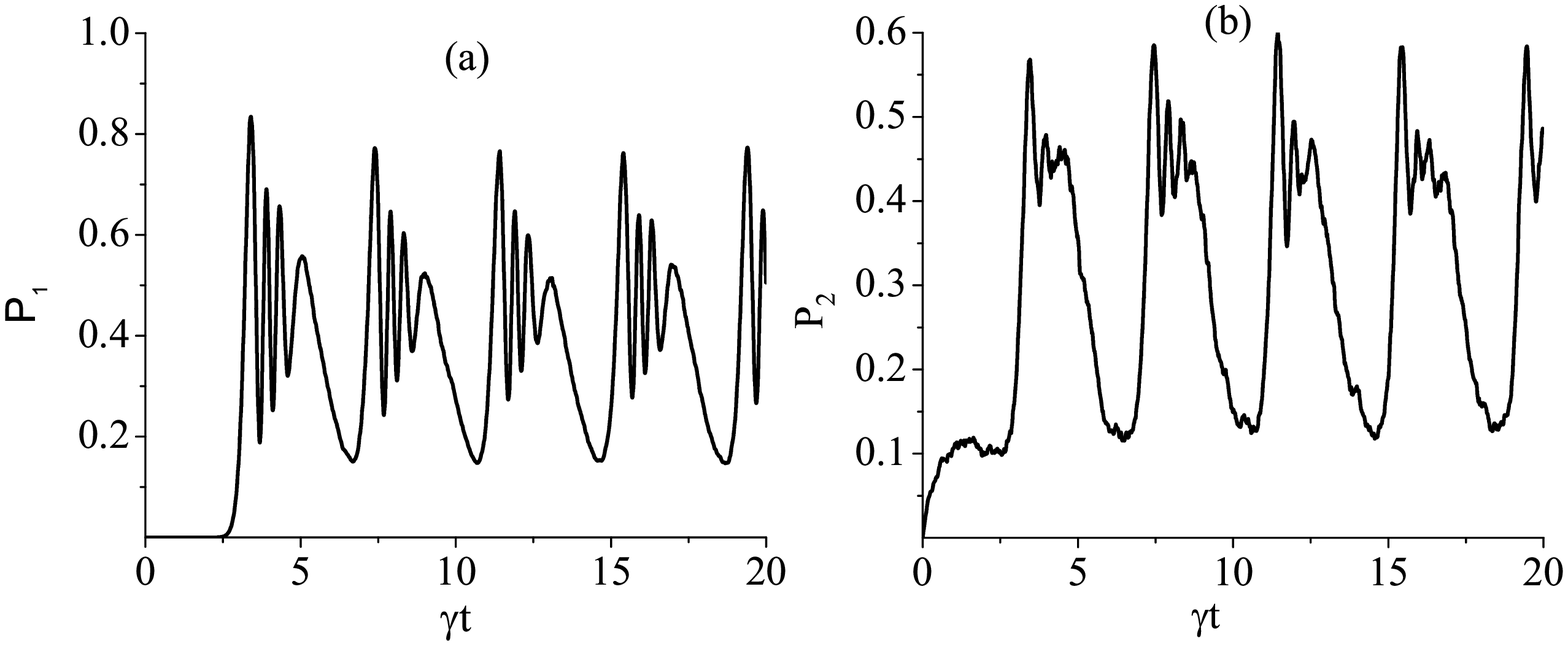}
}
\caption{Time-evolution of the  populations $P_1$ and $P_2$ . The parameters are: (a) $\chi /\gamma = 15$, the maximum amplitude of pump field $\Omega/\gamma =8$ ,  the detuning $\Delta=0$;  (b) $\chi /\gamma = 30$, the maximum amplitude of pump field $\Omega/\gamma =12$ , the detuning $\Delta_{2}=\omega_{0} -\omega_{2}=-\chi$.  
The other parameters are: $\tau = 4{\gamma}^{-1}$, $T=0.8{\gamma}^{-1}.$}
\label{08T}
\end{figure}  
\begin{figure}
\resizebox{0.5\textwidth}{4cm}{%
\includegraphics{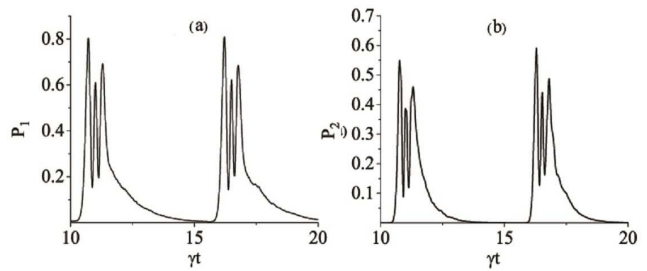}
}
\caption{Time-evolution of the  populations $P_1$ and $P_2$ . The parameters are: (a) $\chi /\gamma = 15$, the maximum amplitude of pump field $\Omega/\gamma =14$ ,  the detuning $\Delta=0$;  (b) $\chi /\gamma = 30$, the maximum amplitude of pump field $\Omega/\gamma =25$ , the detuning $\Delta_{2}=\omega_{0} -\omega_{2}=-\chi$.  
The other parameters are: $\tau = 5.5{\gamma}^{-1}$, $T=0.4{\gamma}^{-1}.$}
\label{pulse_effect}
\end{figure}
For the case when the detuning  $\Delta=\Delta_{1}=0$, i.e. the frequency of driving field $\omega = \omega_{0}$ and only  one-photon transition $|0\rangle\rightarrow|1\rangle$ is effectively realized,  the results for  the population of   $|1\rangle$  Fock
state are shown in Fig. \ref{08T} (a) for the duration of pulses $T=0.8{\gamma}^{-1}$.
  In this case, only two energetic levels of mode are effectively
involved in the Rabi-like oscillations of the populations  $P_0$ and
 $P_1$ of vacuum and first excitation number states  that is
demonstrated in Fig.\ref{08T} (a). Comparing this result with analogous one depicted in Fig. \ref{0}(a) we conclude that the number of Rabi oscillations  increases with increasing of the duration of pulses. The population $P_2$ is shown in Fig. \ref{08T} (b) for the case of two-photon excitation assuming $\Delta=-\chi$. As we see, time-dependence of the population displays Rabi oscillation in difference from the population  shown in Fig.\ref{0} (c) for the case of more short pulses. The analogous situation takes place for photon populations in  increasing of the amplitude of pump field in comparisson with the case of   Fig.  \ref{0}(c). The results for the case of more strong pump field, $\Omega/\gamma =14$ and  $\Omega/\gamma =25$ are presented in Figs. \ref{pulse_effect}. As we see, the Rabi-like oscillations of the populations increase  if the pump amplitude increases. 

\section{Conclusion}\label{Conclusion}

In conclusion, we  have investigated selective excitation of photon-states  in  a lossy Kerr-nonlinear resonator at finite temperatures  driven by a sequence of Gaussian pulses.  In  quantum regime realized for strong Kerr nonlinearities with respect to the rate of damping of
the oscillatory mode one-photon excitation $|0\rangle\rightarrow|1\rangle$ at the frequency $E_{10}=\hbar \omega_0$, as well as two-photon excitation $|0\rangle\rightarrow|2\rangle$ 
at the frequency  $E_{20}=2 \hbar \omega_0 + 2 \chi $ have been analyzed in details. 

We have
demonstrated that the larger photon-number populations of the
resonator can be reached if shaped pulses are implemented. 
The dynamics of mean number of photons and populations of Fock states  follow almost to the train of Gaussian pulses and display Rabi oscillations and the  broadening of the shapes in comparison with Gaussian envelopes. These oscillations  are   due to the radiative decays of the mode in a resonator for time-intervals between pulses. We have  illustrated the increasing of the number of Rabi oscillations with increasing of the duration of pulses as well as with increasing of the amplitude of driving field.

We have demonstrated that temperature effects  leads  to decreasing of  the populations  in comparison with the zero-temperature case on one side and also change cardinally photon statitics of mode on the other side. Particularly, this behaviour on the maximal values of the population $P_1$  in dependence on $n_{th}$  has been illustrated in Fig. 3 for wide range of  the thermal photon level. The quantum statistics of mode is described on the framework of non-stationary second-order correlation function $g^{(2)}$ and the mean photon number fluctuations  $\langle (\Delta n)^2\rangle$.  In this way, we have studied    engineering of photon statistics via a thermal reservoir that is illustrated in Fig. 2(b), (d) and Fig. \ref{05819}. In case of high level of reservoir photons, $g^{(2)}$ describes mainly the   statistics of dissipative mode of the anharmonic oscillator in thermal reservoir with the dips that correspond to the peaks of the averaged photon numbers. The location of these dips showing photon antibunching are determined by time-intervals of Gaussian pulses.

We acknowledge support from the Armenian State Committee of Science, the
Project No.13-1C031. G. Yu. K. acknowledges discussions with Lock Yue Chew,  I. A. Shelykh and S. R. Shahinyan.

\end{document}